\input{epsf.sty}
\documentclass[jmre]{apjrnl}

{\theoremstyle{plain}%

{\theoremstyle{remark}

}
{\theoremstyle{definition}

}

\begin{document}

\title{Hole-burning Diffusion Measurements in High Magnetic Field Gradients}

\author{E. E. Sigmund and W. P. Halperin\\
\additem{Northwestern University, Department of Physics and Astronomy, Evanston, IL,60208}\\
E-mail: esigmund@northwestern.edu}

\date{\today} 

\maketitle

\begin{abstract}

We describe methods for the measurement of translational diffusion in very large static magnetic
field gradients by NMR.  The techniques use a ``hole-burning" sequence that, with the use of fringe
field gradients of 42 T/m, can image diffusion along one dimension on a submicron scale. Two
varieties of this method are demonstrated, including a particularly efficient mode called the
``hole-comb," in which multiple diffusion times comprising an entire diffusive evolution can be
measured within the span of a single detected slice. The advantages and disadvantages of these
methods are discussed, as well as their potential for addressing non-Fickian diffusion, diffusion
in restricted media, and spatially inhomogeneous diffusion.
\end{abstract}



\section{Introduction}

Field-gradient NMR is a widespread research tool.  The common ingredient to any of its applications
is the labelling of space by the Larmor precession frequency of a nuclear species in the presence
of a static magnetic field gradient
$\mathord{\buildrel{\lower3pt\hbox{$\scriptscriptstyle\rightharpoonup$}}\over G}$:

\begin{equation}\label{HB_larmoreq}
\omega _0\left( z \right)=\gamma H_0\left( z \right)=\gamma
\mathord{\buildrel{\lower3pt\hbox{$\scriptscriptstyle\rightharpoonup$}}\over G} \cdot
\mathord{\buildrel{\lower3pt\hbox{$\scriptscriptstyle\rightharpoonup$}}\over r}
\end{equation}
Here $\gamma$ is the gyromagnetic ration of the nucleus, $\omega_0$ is the Larmor frequency, and
$\mathord{\buildrel{\lower3pt\hbox{$\scriptscriptstyle\rightharpoonup$}}\over r}$ is the position
within the sample.  Two applications of this spatial dependence are (1) measurement of structure
(imaging), and (2) measurement of motion (flow and diffusion).  Diffusion measurements have been
performed since the earliest days of NMR research\cite{Hahn_50,CP54,MG58}, and are among the most
widely used image contrast effects in magnetic resonance imaging (MRI).\cite{book-ptcallaghan}

One particular area of interest has been performing NMR experiments in large field gradients.
Recent work on 3-dimensional NMR microimaging with applied gradients of 50 T/m have successfully
achieved voxel volume resolution of 40 femtoliters for biological cell
imaging.\cite{Seeber_00,Ciobanu_02} Another source of large gradients are those in the fringe
fields of NMR magnets\cite{Kimmich91,Sigmund_PRB_01}.  For common superconducting magnets, such
gradients are often in the range of 50 T/m; for high-field resistive magnets ($H_0\approx$ 30 T)
such as at the National High Magnetic Field Laboratory, fringe field gradients greater than 200 T/m
exist. NMR superconducting magnet facilities with Maxwell pair design exist for the purpose of
generating gradients of order 200 T/m\cite{Geil98_CMR}.  Other gradient sources are those outside
of devices designed to probe exterior material, such as the NMR-MOUSE\cite{Eidmann_96}, NMR
well-logging tools\cite{Kleinberg_92,Goelman_95}, or the magnetic resonance force
microscope\cite{Suter_02}. Alternatively, the internal magnetic field gradients within materials
with inhomogeneous magnetic susceptibility have been used to study porous
structure.\cite{Song_Nature_00,Song_PRL_00}  A common issue to many of these cases is that the RF
excitation pulses are ``soft" and do not uniformly excite the sample, either due spatial
inhomogeneity in the $H_1$ field, the limited frequency bandwidth of a finite-duration pulse, or
both.  Recent studies have been performed to fully characterize the spin evolution in CPMG
sequences of many ``soft" pulses to correctly extract diffusive
information\cite{Hurlimann_02,Song_02}. Other analyses have been performed to adapt
multiple-quantum coherence sequences to the field gradient regime\cite{Wiesmath02}.  New techniques
using ``nutation echoes" formed through a combination of inhomogeneities in the static ($H_0$) and
RF($H_1$) fields have been developed\cite{Jershow_98, Sharfenecker_01}, and were included in a
scheme showing the successful recovery of full chemical shift information in the presence of a
static field gradient of 50 mT/m.\cite{Meriles_01,Heise_02} As these studies have shown, prospects
and applications for field-gradient NMR capability are growing.  Thus, it is essential to adapt
existing NMR techniques to the field gradient regime, as well as recognize capabilities that only
large gradients provide.

In this article, we describe methods for the measurement of translational diffusion in large static
field gradients in the fringe field of NMR magnets.  These methods are of the ``hole-burning"
variety, in which long, low power RF pulses are used for spectrally (and thus spatially) selective
irradiations prior to detection.  The time evolution of such ``holes" can be analyzed to extract
diffusion information on a sub-micron scale.  Such a selective excitation technique has been
successfully applied in the past using internal magnetic field gradients to study porous structure
with liquids\cite{Song_PRL_00} and using applied gradients to image the diffusion of
gases\cite{Dimitrov_MRI_99}; the constrast of the present study with that work is in the geometry
and spatial resolution of the resonant slice with a much larger magnitude of the applied gradient.
In addition to providing a viable alternative to dephasing methods, the methods we describe have
potential for deeper study of non-standard diffusion processes, such as anisotropic or restricted
diffusion.

\section{Hole-burning diffusion sequences}

\begin{figure*}
\centerline{\epsfxsize1.0\hsize\epsffile{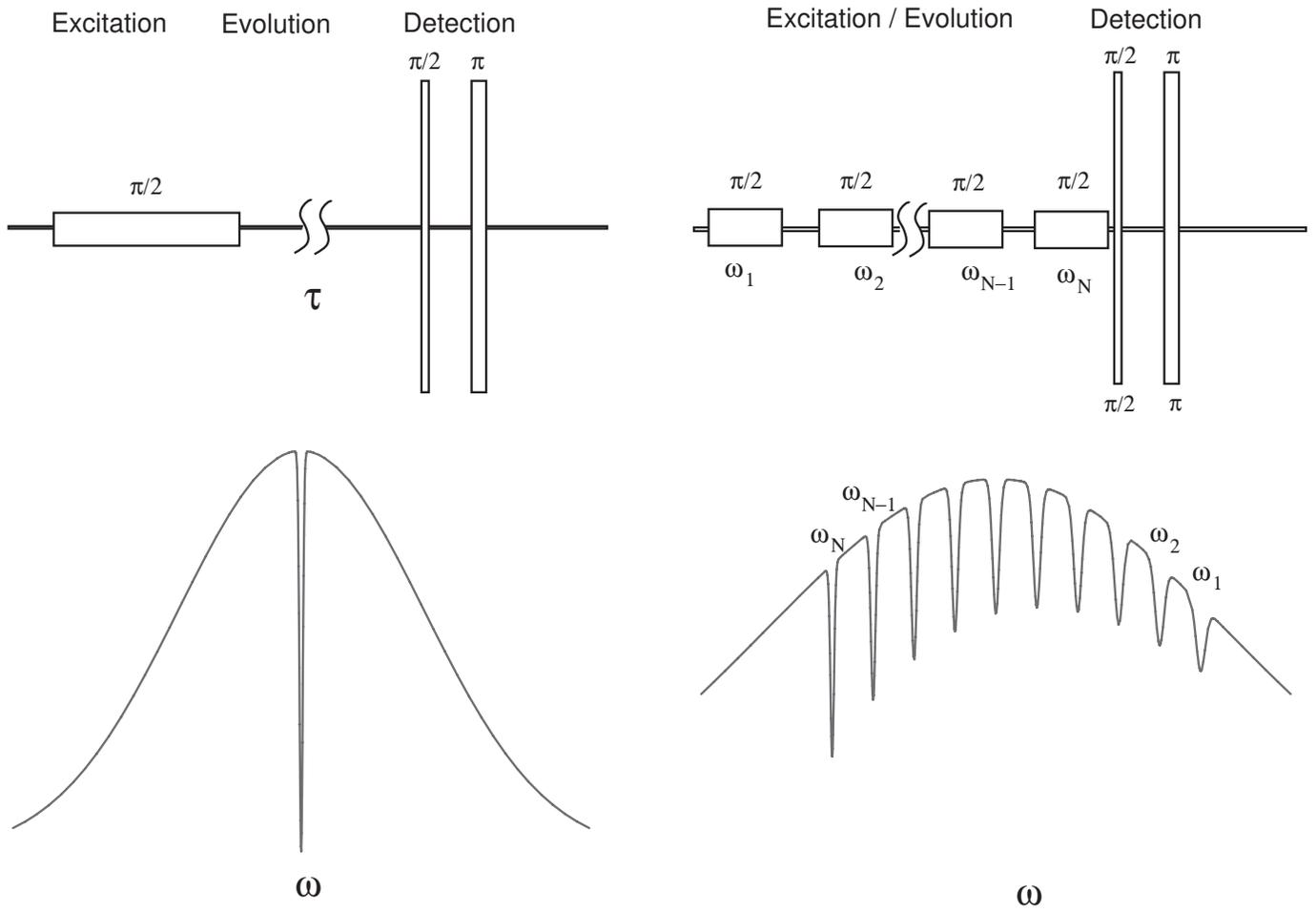}}
\begin{minipage}{1.0\hsize}
\caption{\label{HB_sketches}\small Hole burning diffusion sequences and sketches of spectral
shapes. Left side : single hole burning sequence.  Right side : ``hole-comb" sequence. }
\end{minipage}
\end{figure*}

Two hole-burning sequences are sketched in Fig.\ref{HB_sketches}.  The left panel shows a standard
hole-burning sequence, employed in many NMR experiments as well as other
spectroscopies.\cite{Hunt_68,Duncan_89,Hill_97,Kuhns_82}.  Such an experiment consists of applying
a long, low power pulse to irradiate a narrow band in frequency, given by $\Delta \nu \approx
1/t_p$.  In the presence of a static magnetic field gradient
$\mathord{\buildrel{\lower3pt\hbox{$\scriptscriptstyle\rightharpoonup$}}\over G}$, such a pulse
irradiates a narrow spatial hole perpendicular to the gradient direction.  For example, in a
gradient of $G$=42 T/m, the thickness of a $^{1}$H hole irradiated by a $t_p$=1 ms pulse is
approximately $\Delta z={{\Delta \omega } \over {\gamma G}}={{2\pi } \over {\gamma Gt_p}}\approx$
0.6 $\mu$m. After a diffusion period $\tau$ has elapsed, a broad detection is performed of a large
slice whose thickness is typically a few hundred $\mu$m.  In this case, this takes the form of a
Hahn echo sequence with ``hard" RF pulses.  If $\tau<<T_1$, the spins irradiated with the burn
pulse are ``edited out," and do not appear in the broadly detected signal. As a function of the
evolution time $\tau$, the labelled spins diffuse, widening the hole shape while conserving its
area (in the absence of relaxation).  This time evolution can be analyzed to extract a diffusion
coefficient. This technique works in practice because with large gradients the hole thickness can
be brought to the scale of the diffusion length for an NMR experiment, which is typically a few
$\mu$m for liquids.

In the single-hole sequence, we must wait several spin-lattice relaxation times ($T_1$) after each
acquisition for the magnetization to return to equilibrium.  For long $T_1$, acquiring spectra at
many values of the evolution time $\tau$ is time-consuming.  A variation on the hole-burning
sequence, sketched on the right in Fig. \ref{HB_sketches}, circumvents this inconvenience by using
more of the available detection slice.  In this sequence, not one but a series of hole-burn
excitation pulses are applied.  They are spaced out in time on the scale of the diffusion time, and
each is applied at a different frequency (i.e. position) within the bandwidth of the broad
detection pulse.  After this ``hole-comb", the entire slice is detected, with the result sketched
in the lower right panel of Fig. \ref{HB_sketches}.  The spins in the earlier holes diffuse while
later holes are burned, so that the final spectrum contains a set of snapshots comprising an entire
hole evolution.  This technique assumes a uniform diffusion coefficient and field gradient across
the slice, so all holes broaden at the same rate; for bulk liquids in fringe-field gradients, this
uniformity is excellent (within 0.1\%).  The evolution is acquired in a single transient;
consequently, this sequence has a dramatically improved efficiency, and therefore provides higher
sensitivity. The savings in acquisition time provides a signal-to-noise enhancement of $\sqrt N$,
where $N$ is the number of holes (or evolution times $\tau$) in the sequence.  We note that the
hole-comb measurements we performed were made possible through the fast frequency switching
capability of the MagRes2000$\tiny{^\copyright}$ spectrometer designed by A. P.
Reyes.\cite{Reyes_NHMFL_Rev_01}

\section{Analysis}
In this section we describe briefly the procedures used to extract a diffusion coefficient from a
hole-burning experiment in a fixed field gradient.  We denote the hole-burned absorption spectra at
each evolution time $\tau$ as $F\left( \omega,\tau \right)$ and the unburned spectrum as $F_0\left(
\omega \right)$. The first step is to define the hole shape from the larger slice shape:
\begin{equation}\label{HB_holeiso}
A\left( {\omega ,\tau } \right)=1-{{F\left( {\omega ,\tau } \right)} \over {F_0\left( \omega
\right)}}
\end{equation}
The hole shape $A\left( {\omega ,\tau } \right)$ measures the profile of labelled spins along the
gradient direction as a function of time.  The hole's time evolution with time will be governed by
the diffusive propagator $P\left( {\omega ,\tau ;\omega ',0} \right)$. This gives the probability
that a given spin at frequency $\omega '$ at time $t$=0 will be found at frequency $\omega$ at time
$t=\tau$.  For free, isotropic, diffusion, this 1-D propagator is well known:
\begin{equation}\label{HB_Prop}
P\left( {\omega ,\tau ;\omega ',0} \right)={1 \over {\sqrt {4\pi D\tau }}}e^{-{{\left( {{{\omega
-\omega '} \over {\gamma G}}} \right)^2} \over {4D\tau}}}
\end{equation}
Given the propagator and the initial hole profile $A\left( {\omega ',0} \right)$, the profile at a
later time is found simply by the convolution product,
\begin{equation}\label{HB_convprod}
A\left( {\omega ,\tau } \right)=\smallint P\left( {\omega ,\tau ;\omega ',0} \right)A\left( {\omega
',0} \right)d\omega '.
\end{equation}

For example, we can calculate the explicit form of $A\left( {\omega ,\tau } \right)$ for normal
diffusion and a gaussian hole shape.  Given a gaussian as an initial hole shape, i.e.
\begin{equation}\label{HB_gausshole}
A_g\left( {\omega ,0} \right)=e^{-{{\omega ^2} \over {\sigma _0^2}}}
\end{equation}
Eq. \eqnref{HB_convprod} then becomes
\begin{equation}\label{HB_gauss_eval}
A_g\left( {\omega ,\tau } \right)={1 \over {\sqrt {4\pi D\tau }}}\int_{-\infty }^\infty
{e^{-{{\left( {{{\omega -\omega '} \over {\gamma G}}} \right)^2} \over {4D\tau }}}e^{-{{\omega '^2}
\over {\sigma _0^2}}}d\omega '}
\end{equation}
This gaussian integral can be easily performed by completing the square and using
$\int\limits_{-\infty }^\infty  {e^{-ax^2}dx=\sqrt {{\pi  \over a}}}$.  The result is
\begin{equation}\label{HB_gauss_result}
A_g\left( {\omega ,\tau } \right)={{\sigma _0} \over {\sigma \left( \tau  \right)}}\exp \left(
{-{{\omega ^2} \over {\sigma ^2\left( \tau  \right)}}} \right)
\end{equation}
where
\begin{equation}\label{HB_gauss_sigma}
\sigma ^2\left( \tau  \right)=\sigma _0^2+4\gamma ^2G^2D\tau
\end{equation}
We see that this result conserves the area of the hole, as expected; the amplitude of the gaussian
decays just as the width is enlarged.  Once the initial width has been measured and fixed (through
a spectrum measurement immediately following the burn pulse), all subsequent spectra can each be
fit with a single adjustable parameter $\sigma$.  The resulting square-widths can then be fit to
linear time dependence to extract the diffusivity $D$.  Thus far, in our experiments, we have used
square-wave pulses which would be expected to burn sinc-function rather than gaussian holes.
However, the gaussian is a convenient phenomenological form that, as shown in the experiments
below, accurately describes the broadening of the hole.

The resolution limits of this mode of diffusion measurement are determined by the spin relaxation
times, $T_1$ and $T_2$.  Spin-lattice relaxation causes the tag placed on the burned spins to
evaporate, and fills in the hole without broadening.  In our measurements, we separately measure
the spin-lattice relaxation time $T_1$ and constrain the hole area to decay as $e^{-{\tau
\mathord{\left/ {\vphantom {\tau {T_1}}} \right. \kern-\nulldelimiterspace} {T_1}}}$.  For
sufficiently fast spin-lattice relaxation compared to the diffusion time, the hole broadening due
to diffusion is indetectable.  The minimum hole thickness is another bound on the experiment,
determined either by the diffusivity $D$ (fast limit), or by the spin-spin relaxation time $T_2$
(slow limit).  The hole thickness is controllable only if negligible diffusion or spin dephasing
takes place during the burn pulse length $t_p$.  The combined conditions place the following lower
bounds on measurable diffusivities by this method, depending on the controlling factor in the
minimum hole size (diffusion or spin-spin relaxation)\cite{Sigmund_Thesis_02}.

\noindent For diffusion-limited holes,
\begin{equation}\label{HB_thres_diff}
\left( {{\gamma  \over {2\pi }}} \right)^2G^2D>{1 \over 2}{1 \over {T_1^3}}.
\end{equation}
and for relaxation-limited holes,
\begin{equation}\label{HB_thres_T2}
\left( {{\gamma  \over {2\pi }}} \right)^2G^2D>{1 \over 2}{1 \over {T_1}}{1 \over {T_2^2}}
\end{equation}

Eq. \eqnref{HB_thres_T2} is similar to the limiting diffusivity accessible from stimulated echo
methods\cite{Geil98_CMR,Sigmund_Thesis_02}.  An advantageous application of this method would be to
a system with long $T_2$, such as was found, with $^{1}$H decoupling, in the natural abundance
$^{13}$C signal in glassy glycerol in a previous NMR hole-burning study in a homogeneous
field\cite{Kuhns_82}.

\section{Experiment}

\begin{figure*}
\centerline{\epsfxsize0.5\hsize\epsffile{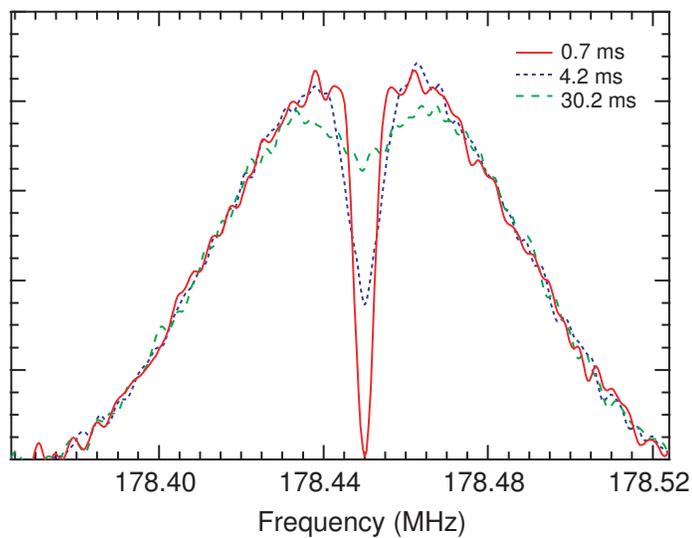}}
\begin{minipage}{1.0\hsize}
\caption{\label{HBfig_HB_exp}\small Hole-burning diffusion measurement by $^{1}$H NMR in propylene
at T = 295 K in a gradient of $G$ = 42 T/m.}
\end{minipage}
\end{figure*}

\begin{figure*}
\centerline{\epsfxsize1.0\hsize\epsffile{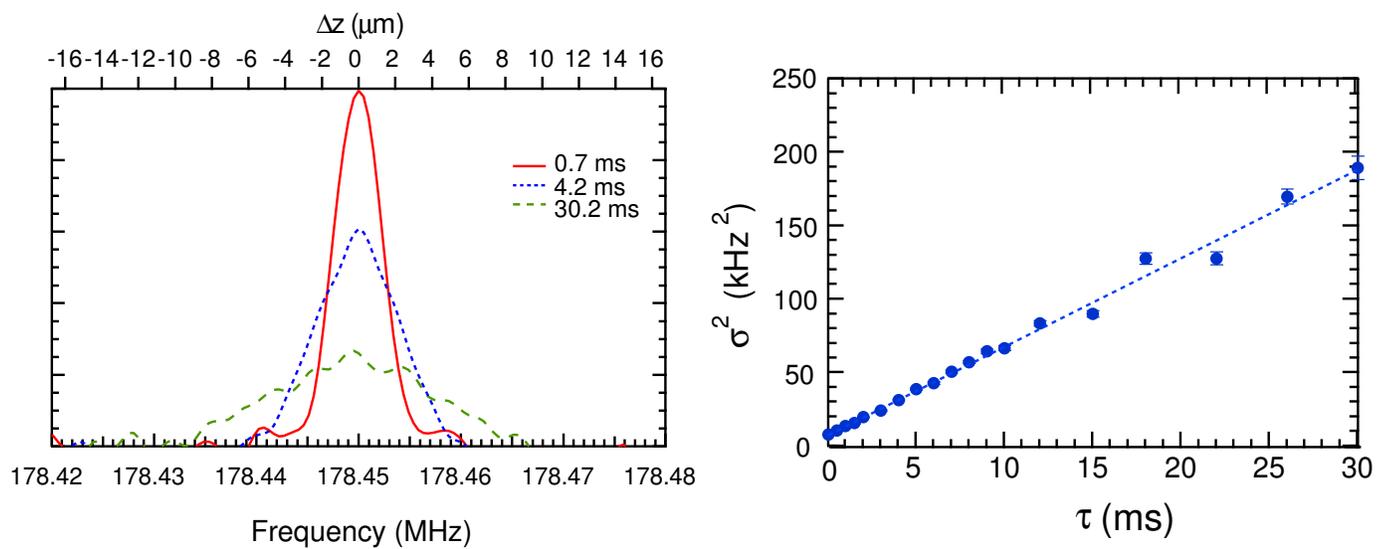}}
\begin{minipage}{1.0\hsize}
\caption{\label{HBfig_HB_fits}\small Fits of hole-burning diffusion measurement by $^{1}$H NMR in
propylene carbonate at T = 295 K in a gradient of $G$ = 42 T/m.}
\end{minipage}
\end{figure*}

Fig. \ref{HBfig_HB_exp} shows a measurement by a $^{1}$H NMR hole-burning sequence of the
diffusivity of propylene carbonate at $T$=295 K.  The corresponding individual hole shapes and
their gaussian fits are shown in Fig. \ref{HBfig_HB_fits}.  The spin-lattice relaxation time $T_1$
in this case is of the order of seconds, much longer than the evolution times of the experiment (up
to 30 ms).  The resulting diffusivity is $D$=(4.8 $\pm$ 0.1) 10$^{-6}$cm$^2$/s, which compares well
with a diffusion measurement by a more standard stimulated echo dephasing measurement with the same
sample and gradient.

\begin{figure*}
\centerline{\epsfxsize0.5\hsize\epsffile{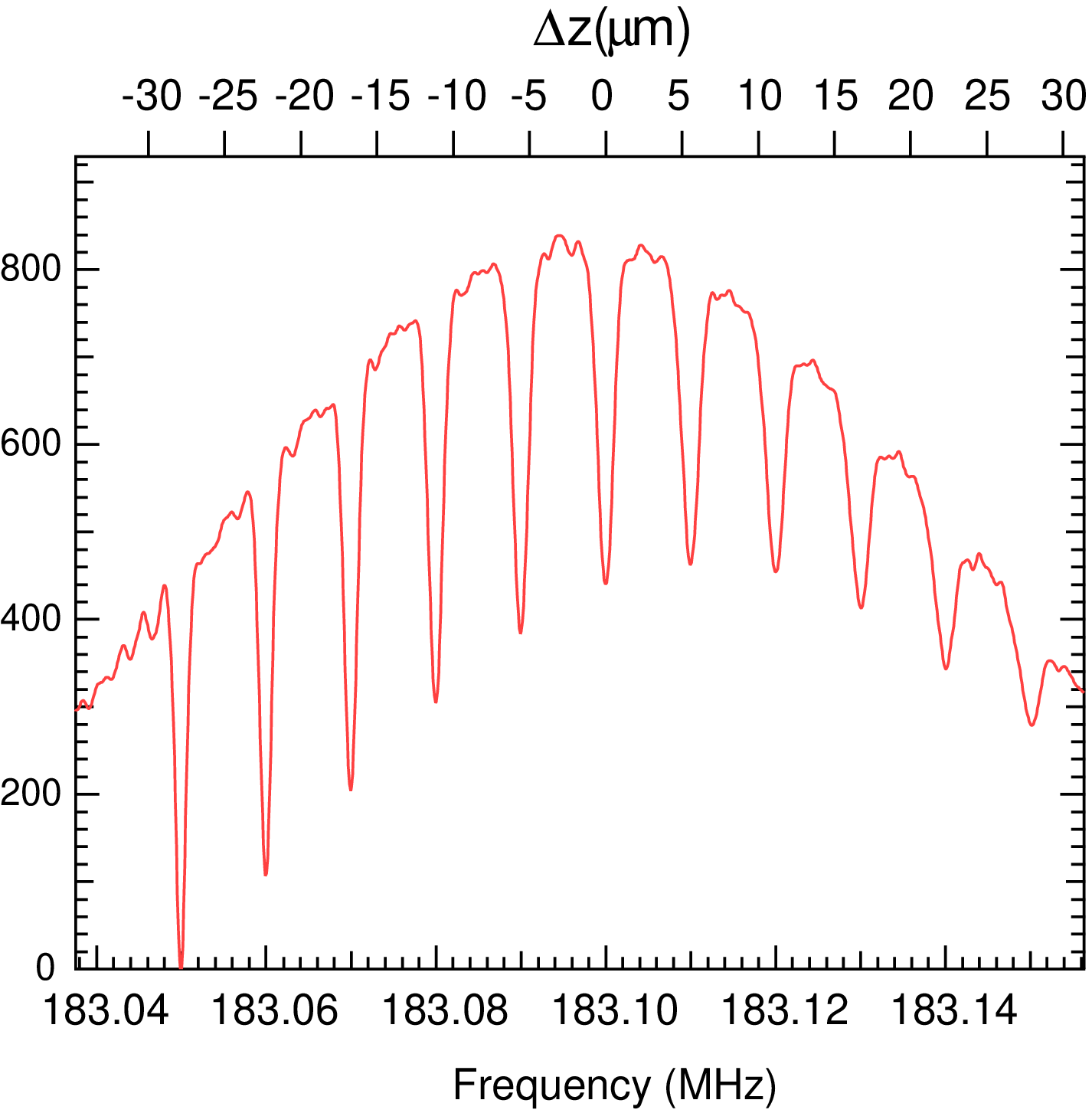}}
\begin{minipage}{1.0\hsize}
\caption{\label{HBfig_HBC_exp}\small Hole-comb diffusion measurement by $^{1}$H NMR in glycerol at
T = 296 K in a gradient of $G$ = 42 T/m.}
\end{minipage}
\end{figure*}

\begin{figure*}
\centerline{\epsfxsize0.7\hsize\epsffile{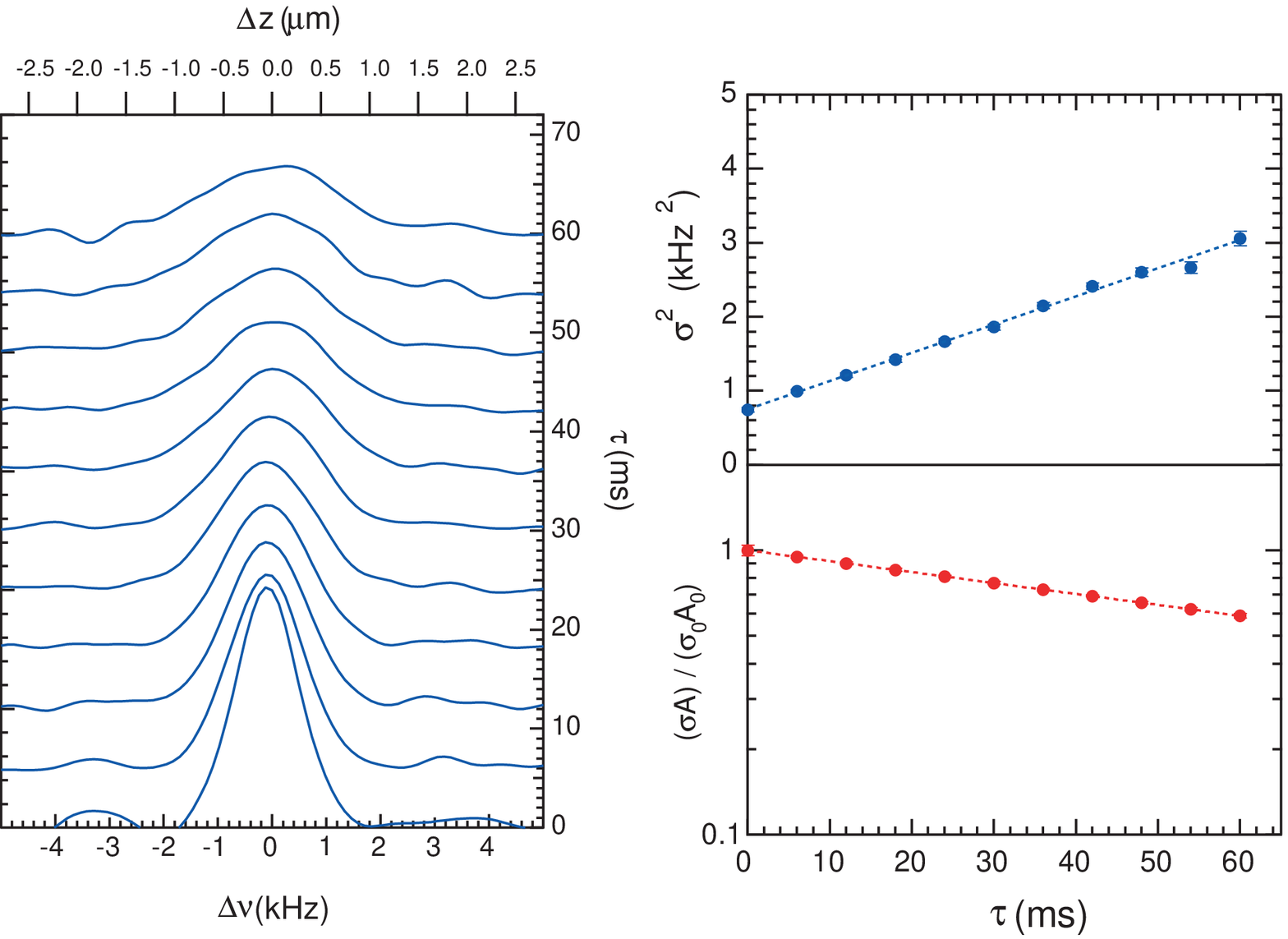}}
\begin{minipage}{1.0\hsize}
\caption{\label{HBfig_HBC_fits}\small Fits of hole-comb diffusion measurement by $^{1}$H NMR in
glycerol-$^{13}$C$_2$ at T = 296 K in a gradient of $G$ = 42 T/m.}
\end{minipage}
\end{figure*}

Fig. \ref{HBfig_HBC_exp} shows a measurement by a $^{1}$H NMR hole-comb sequence of the diffusivity
of glycerol-$^{13}$C$_2$ at $T$=296 K.  The corresponding individual hole shapes and their gaussian
fits are shown in Fig. \ref{HBfig_HBC_fits}.  In these fits, the area of the hole was constrained
to decay as $e^{-{\tau  \over {T_1}}}$, with a spin-lattice relaxation time of $T_1$=112 ms
measured separately by a standard saturation recovery sequence.  The resulting diffusivity is
$D$=(2.86 $\pm$ 0.05) 10$^{-8}$cm$^2$/s, which again compares well with a stimulated echo diffusion
measurement.

\begin{figure*}
\centerline{\epsfxsize0.5\hsize\epsffile{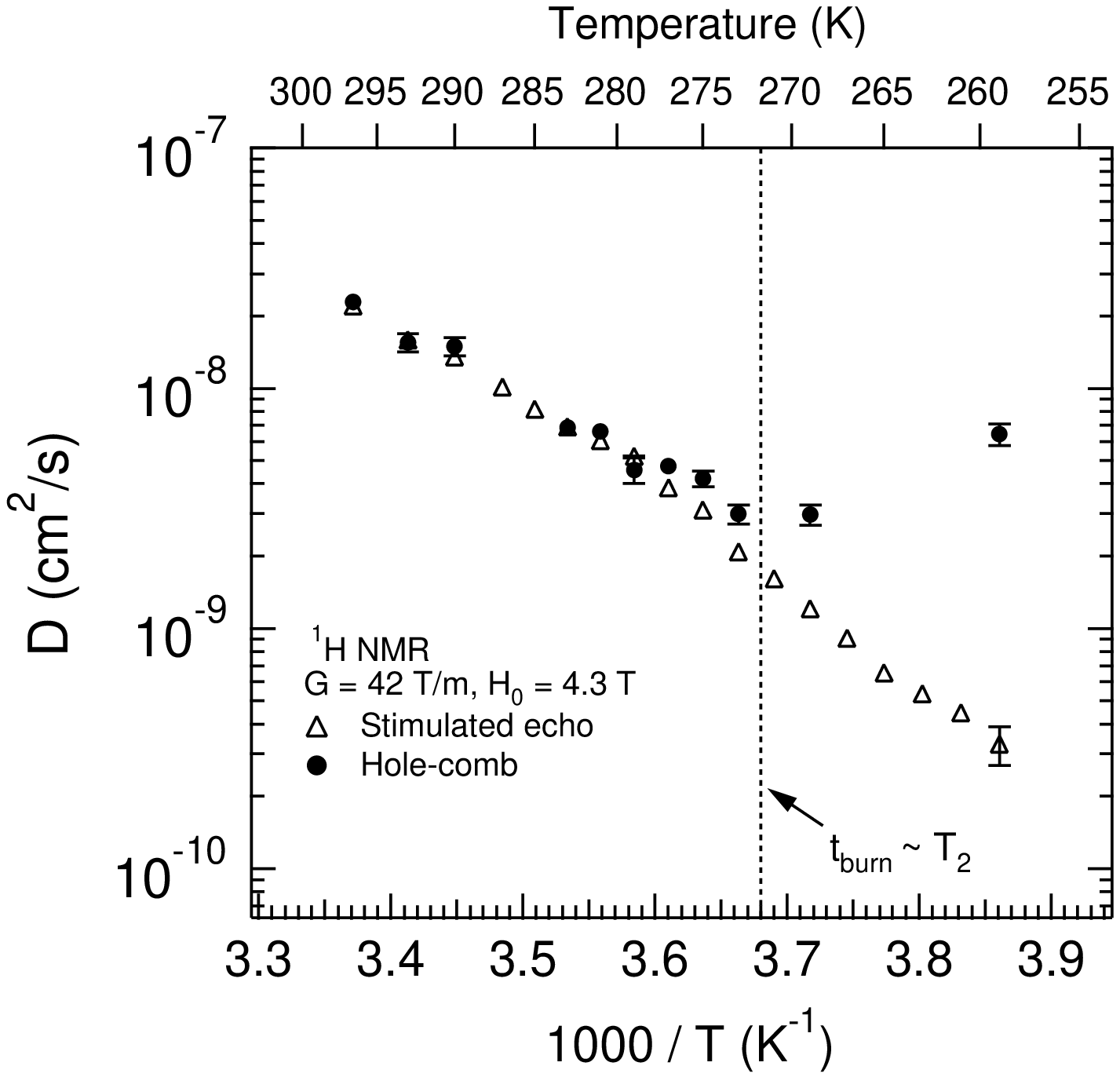}}
\begin{minipage}{1.0\hsize}
\caption{\label{HBfig_HBC_v_SE}\small $^{1}$H NMR comparison of diffusivity measurements by
stimulated echo and hole-comb techniques in the same sample of glycerol-$^{13}$C$_2$, with an
applied gradient of $G$=42 T/m.  The two techniques agree until the spin-spin relaxation time $T_2$
is of order the burn pulse width $t_{burn}$; below this temperature translational diffusion cannot
be correctly discerned by the hole-comb technique. The error bars are computed from the statistical
accuracy of the fits, and are not shown if smaller than the symbol size.}
\end{minipage}
\end{figure*}

Finally, we show in Fig. \ref{HBfig_HBC_v_SE} a comparison of temperature dependences of the
glycerol-$^{13}$C$_2$ $^{1}$H diffusion coefficient measured by two different methods: stimulated
echo and hole-comb measurements.  The agreement is good for higher temperatures, until the point
that the burn pulse interval ($\approx$ 1 ms) approaches the spin-spin relaxation time, $T_2$.  At
this point the hole-comb sequence becomes unreliable.

\section{Applications}
For normal diffusion, a convenient form can be derived for the hole evolution given any initial
hole shape.  Alternatively, if the diffusive propagator is non-Fickian, this measurement can serve
to map out its behavior by convoluting with a known excitation function.  This capability
distinguishes the hole-burning sequence from the dephasing sequences.  Given a spectral evolution
profile from a hole-burning sequence, higher order moments can be calculated which provide more
information on the propagator.  Specifically, so long as the propagator is stationary, i.e. depends
only on the difference of the observation times $\tau=t-t'$ and positions $Z=z-z'$, it can be shown
that any spatial moment of an evolved hole spectrum $A\left( {z,\tau} \right)$, defined by

\begin{equation}\label{HB_Moments_def}
M_z^n\left[ {A\left( {z,\tau } \right)} \right]\equiv {{\int {dz\left( {z-\left\langle z
\right\rangle } \right)^nA\left( {z,\tau } \right)}} \over {\int {dzA\left( {z,\tau } \right)}}}
\end{equation}

\noindent is simply the sum of the initial hole moment and that of the diffusive propagator:

\begin{equation}\label{HB_Moments_eq}
M_z^n\left[ {A\left( {z,\tau } \right)} \right]=M_z^n\left[ {A\left( {z,0} \right)}
\right]+M_Z^n\left[ {P\left( {Z,\tau } \right)} \right]
\end{equation}

Thus, to the extent that any function can be reconstructed with knowledge of its moments, this
provides a method for determining a non-Fickian propagator's spatial dependence.

The hole-burning sequence is potentially useful in the study of porous media; if the initial hole
size can be made less than the smallest confinement length scale, the crossover from free to
restricted diffusion can be observed by steadily increasing the initial hole size.  Such a
variation is similar to that accomplished by variation of the first interval in a stimulated echo
sequence. Finally, the spatial resolution of the hole-burning sequence is well-suited to problems
of transport near surfaces; the diffusion coefficient in a liquid can be inspected on a sub-micron
scale at arbitrary distances from a solid-liquid or solid-gas interface.

\section{Conclusions}

We have demonstrated a hole-burning NMR diffusometry technique in large magnetic field gradients
($G$ = 42 T/m).  The advantages this technique offers include spatial resolution and a higher
descriptive capability for non-standard diffusion behavior.  Future directions include application
to porous media and diffusion near surfaces on the sub-micron scale.

\begin{acknowledge}
We thank Y.-Q. Song and S. Lee for useful comments.  This work has been supported by the NSF-MRSEC
at the Materials Research Center at Northwestern University, grant \# DMR-0076097.
\end{acknowledge}

\addcontentsline{toc}{chapter}{References}
\renewcommand{\baselinestretch}{1}  
\small\normalsize      
\bibliographystyle{vemi_bib_sty}
\bibliography{HoleBurningDiffusion}


\end{multicols}
\end{document}